# Implementation of Energy Efficient Single Flux Quantum (eSFQ) Digital Circuits with sub-aJ/bit Operation

M H Volkmann[1], A Sahu[2], C J Fourie[1], and O A Mukhanov[2]

[1] Department of Electrical and Electronic Engineering, Stellenbosch University, Private Bag X1, 7602, Stellenbosch, South Africa

[2] HYPRES, Inc., 175 Clearbrook Rd, Elmsford, NY 10523, USA

E-mail: 15164187@sun.ac.za

**Abstract**. We report the first experimental demonstration of recently proposed energy-efficient single flux quantum logic, eSFQ. This logic can represent the next generation of RSFQ logic eliminating dominant static power dissipation associated with a dc bias current distribution and providing over two orders of magnitude efficiency improvement over conventional RSFQ logic. We further demonstrate that the introduction of passive phase shifters allows the reduction of dynamic power dissipation by about 20%, reaching ~0.8 aJ per bit operation. Two types of demonstration eSFQ circuits, shift registers and demultiplexers (deserializers), were implemented using the standard HYPRES 4.5 kA/cm$^2$ fabrication process. In this paper, we present eSFQ circuit design and demonstrate the viability and performance metrics of eSFQ circuits through simulations and experimental testing.

## 1. Introduction

Continuous improvement in computing fuelled by scaling of underlying digital and memory devices has come under question as their physical structures approach atomic dimensions. It is no longer expected that simple scaling of existing silicon-based devices and multi-core architectures will lead to the next level in high-end computing. The dissipated power became the new fundamental limiting factor defining processor performance [1, 2]. On-going concentration of computing resources in data centres for supercomputers and cloud computing actuated the need for improvement in energy-efficiency of computing technologies, including the core digital and memory circuits [3, 4]. Energy-efficiency has become the dominant metric for the next generation of computing technologies [1].

Further scaling of Si CMOS (complimentary-metal-oxide-semiconductor) devices is not expected to achieve the required energy-efficiency fast enough to meet the demands for the next generations of high-end computing systems [1]. In addition to the energy consumed during CMOS gate switching, the energy used for digital data movement is even more significant and difficult to scale. To make things harder, the energy efficiency requirements preclude the use of many microprocessor design innovations developed over past 20 years [1]. All these open up an opportunity to explore alternative new types of devices and materials with inherently higher potential for high energy efficiency, and thereby ultimately leading to the development of energy-proportional computers that consume energy proportional to the activity level – no energy use while idling and gradually increasing consumption with increasing work load [5].

Superconducting single flux quantum (SFQ) Josephson devices with fast (~1 ps) and low-energy (~$10^{-19}$ J) switching coupled with fast and lossless interconnects has been viewed as a potential alternative technology for high-end computing [4, 6]. Relatively mature Rapid Single Flux Quantum (RSFQ) [7] superconducting cryogenic technology is already being used for practical implementations of cryocooled direct digital receivers for satellite communications and signal intelligence operating at clock frequencies of tens of gigahertz [8-11].

Superconducting RSFQ technology is exploiting quantized magnetic flux to encode clock and digital data as SFQ voltage pulses with area $\Phi_0 = h/2e$ ~$2.06 \cdot 10^{-15}$ *Wb*. The switching energy associated with SFQ pulses crossing Josephson junction is of the order of $I_c \Phi_0$ ~$10^{-19}$ *J* for typical junction critical current $I_c$ ~ 0.1 *mA* determined by thermal noise for a 4 K operation. Low loss and low dispersion superconducting microstrip lines enable ballistic transfer of the SFQ picosecond signals without signal amplification on chip and between chips with speeds of the order of speed of light [12-15]. The quantized nature of SFQ voltage pulses $V(t)$, $\int V(t) \, dt = \Phi_0$, ensures high tolerance to circuit parameter variability. All these features: the gate energy set by thermal noise rather than device scaling, the energy-free ballistic interconnect rather than the data movement energy proportional to the interconnect length, circumvent key problems CMOS faces today.

However, the static power dissipation of conventional RSFQ circuits is significantly larger (by ~ 100x) than their switching power. It is associated with a resistor-based bias distribution network to deliver the required dc bias current for each gate regardless of circuit operation load. This makes RSFQ circuits difficult to extend to a CMOS-level integration density required for processors and contradicts the recent emphasis on energy efficient, energy-proportional computing.

The latest developments in SFQ circuits have been focused on addressing high static power dissipation including reduced static power versions of RSFQ [16-18], ac-powered reciprocal quantum logic (RQL) [19-20], and new energy-efficient generations of RSFQ with zero-static power: ERSFQ and eSFQ [21-23]. Most of these new circuit approaches were tried and various test circuits demonstrated. The eSFQ approach was only discussed theoretically [21].

In this paper we report the first experimental demonstration of eSFQ circuits including shift registers and demultiplexers. These eSFQ circuits make use of superconducting dc bias current dividers and thus avoid static power dissipation. Until recently, this was considered impossible in RSFQ-type circuits, since it would lead to superconducting phase and average voltage imbalances caused by data SFQ propagation in superconducting Josephson circuits. In eSFQ circuits, all RSFQ core advantages of high-speed, dc power, internal memory, local clock control along with the already developed RSFQ circuit designs are largely preserved. The elimination of static power dissipation in eSFQ circuits results in over two orders of magnitude reduction of overall circuit power as compared to conventional RSFQ circuits.

We also report on our attempts to reduce dynamic power dissipation of SFQ gate. This is achieved by employing passive superconducting phase shifters resulting in reduction of circuit bias current and, thus, dynamic power dissipation. Furthermore, since the eSFQ circuit dc bias is controlled by the SFQ clock, one can manage dynamic power by managing the distribution of SFQ in the clock network. It is possible to turn off the SFQ clock for a particular part of a processor and effectively stop the circuit operation, i.e. achieve zero dynamic power, while maintaining the internal state of the affected circuit. This feature is compatible with the coveted goal of achieving energy-proportional computing as the ultimate energy-efficient machine [5].

## 2. Principles of eSFQ logic
*2.1. Power dissipation in RSFQ logic*
In SFQ circuits including all types of RSFQ and RQL, digital information is encoded, processed and transported with single flux quanta or SFQs. Physically, SFQ circuits comprise a network of active Josephson junctions, and passive lossless and low dispersion superconducting strips forming inductors and microstrip lines. The energy is spent when an SFQ traverses a Josephson junction, causing a discrete $2\pi$ superconducting phase slip. This process is equivalent to the regeneration [24] of a quantized voltage pulse $V(t)$. The energy dissipated in the shunt resistor during this event requires replenishment for continuous circuit operation. It is fed into the circuit by means of current biasing.

The simplest way towards achieving this is to employ a dc current bias, as is the case for conventional RSFQ logic introduced in 1985 [25, 26]. Resistive dividers are employed to distribute current to the different bias points in a circuit, as depicted in figure 1.

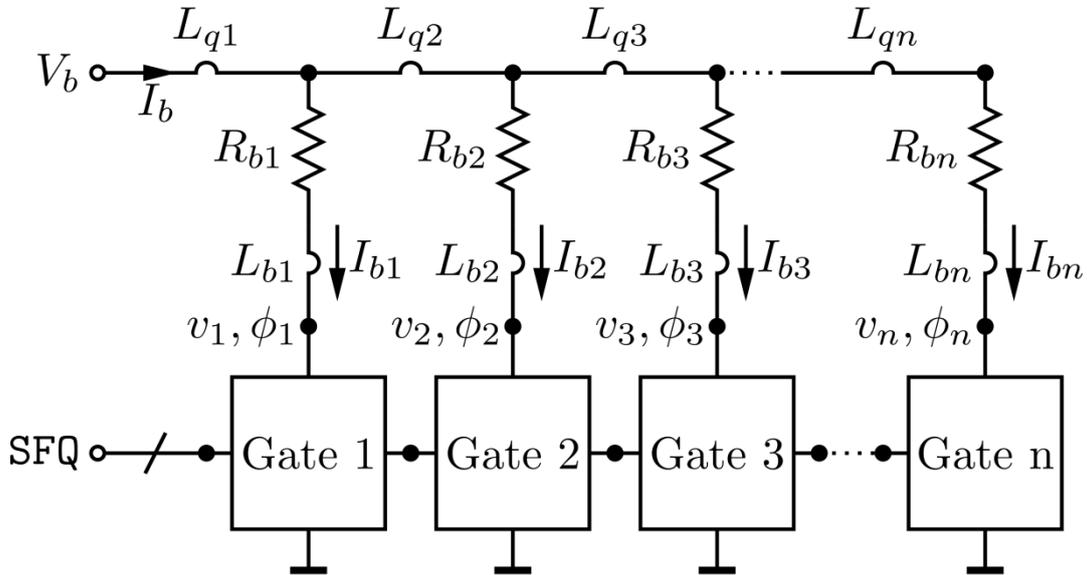

**Figure 1**. Conventional RSFQ biasing. The total dc bias current required by the RSFQ gates is injected at the node labeled $V_b$. When the RSFQ logic gates are in the quiescent state, the voltage at the bias injection points is $v_1, v_2, \ldots = 0$. Resistors $R_{b1}, R_{b2},\ldots$ divide the bias current to allocate each gate the correct proportion. Parasitic bias-path inductance does not influence the final division of the dc current.

To preserve logical integrity of the gates, the bias may not disturb the phases within the circuit. An RSFQ logic relies on the assumption that, during circuit operation (after bias-current ramp-up), superconducting phase at circuit nodes is not altered by means other than passing SFQ pulses. There are exceptions to this for interface circuits, for example in cells such as the DC-resettable latch [27] and SFQ/dc converter [28], but the concept remains valid.

This resistive approach to biasing is stable, as long as the bias voltage $V_b$ exceeds the maximum voltage at the bias injection points significantly. If this is not the case, then the voltage drop over the corresponding bias resistor may no longer be close to $V_b$ at all times, skewing the bias-current allocation. The maximum average voltage $v_i$ at any bias injection point is $(v_i)_{max} = \Phi_0 f$, where $f$ is the (highest) clock frequency of the logic circuit. That is,

$$v_i \leq \Phi_0 f \quad \forall i. \tag{1}$$

Given this discussion, power dissipation in conventional RSFQ logic circuits can be neatly classified into two categories: *static power dissipation* and *dynamic power dissipation*. Static power dissipation occurs in the bias network and is independent of logic switching. Dynamic power dissipation occurs in the logic circuitry as a result of SFQ switching events. Since $V_b \gg v_i$ for all gates $i$, the static power dissipation for a circuit of $n$ gates is given by

$$P_s \approx V_b \cdot I_b, \tag{2}$$

and the dynamic power dissipation is

$$P_d = \sum_{k=1}^{n} v_k \cdot I_{bk} \leq \Phi_0 f \cdot I_b. \tag{3}$$

The ratio between static power dissipation and dynamic power dissipation is thus

$$\frac{P_s}{P_d} \geq \frac{V_b}{\Phi_0 f}. \tag{4}$$

For typical bias voltages employed by RSFQ logic ($V_b \sim 2.6$ mV) and for typical clock frequencies (currently tens of GHz), this ratio ranges between one and two orders of magnitude. Clearly, to achieve ultra-low-power operation, addressing the static power dissipation is paramount.

A straightforward approach to reduce the dominant static power dissipation would be to reduce $V_b$ as much as possible. However, bringing $V_b$ closer to $\Phi_0 f$ makes the bias current distribution less stable, requiring a corresponding increase in $L_b$. This LR-biasing approach was first introduced in 1997 [29] and lately was further elaborated [16-18]. While at first glance this method may seem to support an arbitrarily low static power reduction, the $L/R$ time constant of the branch imposes an upper bound on the frequency of operation, decreasing the effectiveness and applicability of LR-biasing.

*2.2. Eliminating static power dissipation*
Recently, alternative approaches have emerged that are capable of eliminating static power dissipation of conventional RSFQ circuits: RQL [19], as well as energy-efficient ERSFQ and eSFQ logic [21, 22]. Both of them get rid of the resistor-based dc bias current distribution network solely responsible for dominant static power dissipation. In RQL, this is achieved by shifting from a dc bias current to an ac bias current delivered via a special superconductor microwave distribution network and transformers integrated to each gate. In ERSFQ and eSFQ, the dc biasing is preserved, while bias resistors are replaced with Josephson junctions limiting dc bias current to gates [21, 22].

An examination of the I-V curve of the Josephson junction (see figure 2(a)) leads to the observation that, at the transition point between the two regions of operation, the junction transmits a current $i = I_c$ at zero voltage. This suggests that the junction may be employed as a limiting device to set up the required bias current level for a gate.

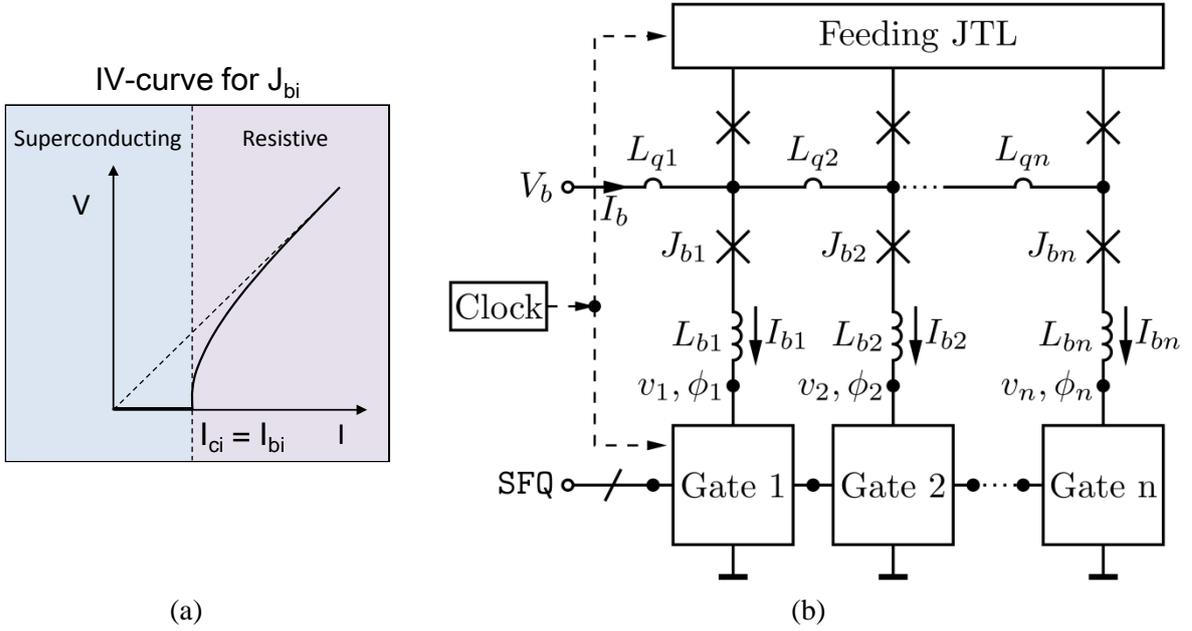

(a)                                          (b)

**Figure 2.** The current-limiting properties of the Josephson junction (a) can be exploited to achieve the desired bias current distribution (b).

Injecting a total bias current $I_b = \sum_i I_{ci}$ forces each junction $J_{bi}$ to transmit its critical current, which is designed to the desired $I_{bi}$ of each branch:

$$I_{ci} = I_{bi} \; \forall i. \tag{5}$$

This enables current limiting with zero voltage drop across the junction, essentially achieving $P_s = 0$. However, when the biased gates are active (switching), the voltage at the bias injection points $v_i$ is non-zero. To maintain the correct region of operation of the current-limiting junctions, the voltage on the bias line must at least equal the voltage at the bias injection points. In ERSFQ circuits, this can be achieved by connecting a so-called *feeding Josephson transmission line (JTL)* to the bias line as depicted in figure 2. If the feeding JTL is connected to the same clock as the fed gates, the bias injection points on the JTL experience a $2\pi$ phase slip at every clock period. This ensures that the voltage on the bias line is kept at $v \approx \Phi_0 f_{\text{clk}}$, which we know from (1) to match or exceed the voltage at all other bias injection points. Hence, the voltage across the biasing branches is kept at or just above zero, maintaining the correct operational region of the limiting junctions.

Although the average current transmitted by the limiting junction is equal to its critical current, the instantaneous current may deviate when the gates are active. To reduce this effect, a limiting inductor $L_b$ must be employed. For typical required bias currents (several 100 µA), guaranteeing a deviation of $\Delta I_b < 5\%$ requires a (comparatively large) inductance of $L_b \approx 400$ pH per branch. This translates to a larger area required for ERSFQ gates as compared to conventional RSFQ circuits. Nonetheless, complex ERSFQ circuits have been successfully demonstrated [22, 23].

*2.3. eSFQ Circuits*

While ERSFQ successfully minimizes static power dissipation with minimal modification to existing RSFQ gates, the addition of large averaging inductors and feeding JTLs brings about certain design challenges. These inconvenient requirements can be removed in the recently proposed eSFQ logic [21]. The eSFQ bias network (figure 3) is topologically similar to the ERSFQ bias network. The chief difference lies in the size of the limiting inductor $L_b$ and the absence of a feeding JTL. Instead, the function of the feeding JTL is accomplished by the SFQ clock distribution network, which is largely a JTL network.

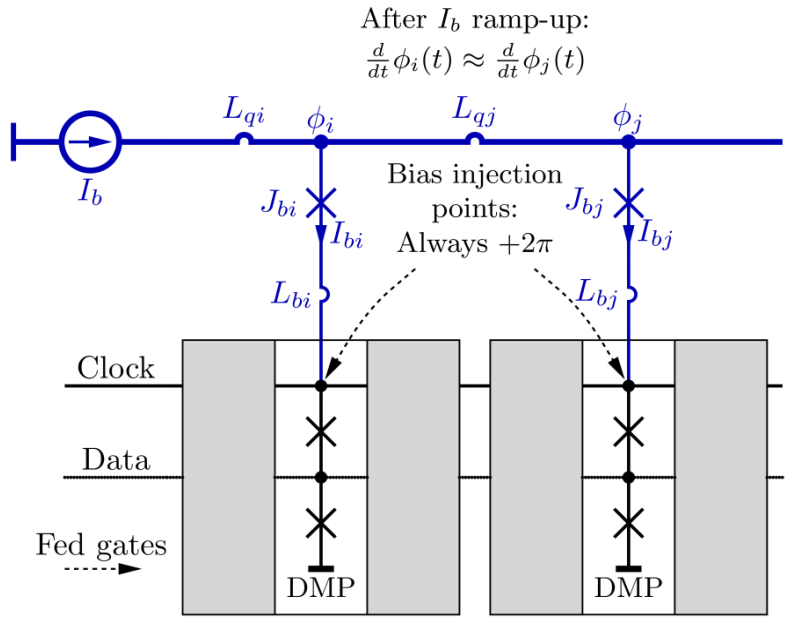

**Figure 3.** eSFQ biasing principle: Exploiting the current-limiting properties of the Josephson junction without the need for a large $L_b$ is possible when the phase gradients between different bias terminals remain constant ($\dot{\phi}_i(t) \approx \dot{\phi}_j(t) \; \forall i,j$). This can be achieved by ensuring that all bias terminals are connected to the clock net, as each point on the clock net experiences the same phase shift ($+2\pi$) during each clock period. Most RSFQ gates contain a decision-making pair (DMP) which is interrogated by the clock signal, making for a suitable bias terminal (shaded gate area represents other gate circuitry).

To enable biasing in this manner, care must be taken to ensure that the limiting junctions do not switch during circuit operation. This obviates the need for a large averaging inductor $L_b$. To ensure this, the junction phases must be kept at sub-critical level during circuit operation. This requirement is met trivially while the fed circuitry is quiescent, as the phase at each bias injection point then remains constant. Therefore, the phase across the limiting junctions, once established during ramp-up of the bias current, remains undisturbed (at sub-critical level).

When the fed circuitry is active, however, the phase at the bias injection points does not remain constant. Switching events in the fed circuitry bring with it phase increments of $2\pi$. The key observation here is that for a general RSFQ-type circuit, these phase increments may or may not occur at the bias terminals, depending on data SFQ propagation. So, usually, in any given clock period, the phase at a bias injection point either jumps by $2\pi$, or remains constant. Since data are generally random, different bias injection terminals accumulate different total phase, creating a changing phase difference between terminals over time. This phase imbalance is acceptable in conventional RSFQ and LR-biased RSFQ, in which the normal metal bias resistors of biasing network leak out the accumulated magnetic flux and keep phases decoupled. In the case of an all-superconducting biasing network, were these to remain unequalized, a parasitic supercurrent would flow across the phase difference, skewing the bias current distribution. In ERSFQ circuits, phase equalization is achieved by compensating phase increments occurring once the parasitic current increases the bias of a gate beyond the critical current of the gate limiting bias junction. In all these different RSFQ circuit approaches, the phase imbalances equalize in an asynchronous manner, which adds some degree of uncontrollable variations to dc bias currents. These contribute to the time jitter and limit ultimate circuit performance.

In contrast, the eSFQ approach removes this limitation by eliminating the very source of phase imbalance – the data dependent $2\pi$ phase increments at the bias injection terminals. In eSFQ, the biasing network is designed so that in each period, the phase at *all* bias injection terminals goes through the same change (figure 3). In this way, the phase across each limiting junction does not change after the initial ramp-up of bias currents, whether the active circuitry is switching or not. Although conventional RSFQ gates are not designed with this requirement in mind, many RSFQ cells lend themselves well to conversion to eSFQ as most RSFQ cells are clocked, requiring that a $2\pi$ phase increment occur somewhere in the cell during each period. The next section relates our efforts at achieving this for several common cells.

**3. Design of eSFQ demonstrator circuits**

Most RSFQ circuits are generally well-suited to conversion to eSFQ. Consider, for example, a JTL transmitting the clock signal. As each junction in the JTL switches during each clock period (experiencing a $2\pi$ phase slip), bias current injection can occur at arbitrary locations. RSFQ logic gates rely on serially connected pairs of Josephson junctions – Decision-Making Pairs (DMPs) – to perform a function and control output. In a DMP, exactly one junction switches whenever an interrogating SFQ pulse is applied. Whichever junction switches, the sum of the phases across the DMP increases by $2\pi$. When the interrogating pulse is a clock signal, the phase atop the DMP meets the requirements of an eSFQ bias injection terminal.

D-cell (D flip-flop) conversion from RSFQ to eSFQ is depicted in figure 4. Conventionally, the D-cell is biased so that it initially stores a logic "0". Such biasing occurs on junction $J_1$, which only experiences a $2\pi$ phase increment when a pulse arrives at *In*. However, exactly one pulse arrives at *Clock* during every clock period, ensuring a $2\pi$ phase increment across the DMP. Hence, for eSFQ, the bias injection point is moved to the DMP. As a side effect, the converted D-cell stores a logic "1" after initial bias ramp-up.

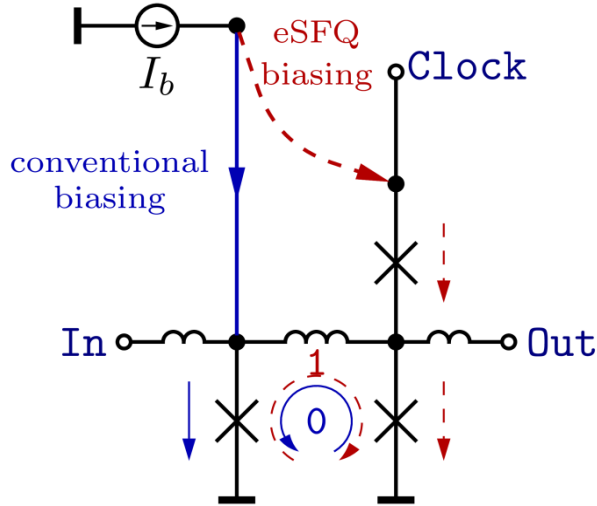

**Figure 4.** Conversion of standard RSFQ D flip-flop to eSFQ.

In an analogous manner, most RSFQ cells can be converted to eSFQ. Note that certain RSFQ cells do not lend themselves as intuitively to conversion to eSFQ, such as the T flip-flop (TFF), which is not clocked and thus, in conventional form, does not have a bias injection point suitable to eSFQ. In this case, their design should be modified to the unbiased, "supply-free" design to enable operation in a ballistic mode. SFQs enter such cells ballistically from biased adjacent cells. For example, the known unbiased version of the TFF cell [30] may be sandwiched between two storage elements, so that SFQ pulses will enter and leave the TFF cell ballistically.

We chose two circuits for the first experimental demonstration of eSFQ approach: a shift register and a demultiplexer (deserializer). A shift register is a typical benchmark circuit used on assessment of a new circuit technology. It is also widely used in digital and mixed-signal circuits. A demultiplexer is a circuit of a considerable practical significance, e.g., for use with superconducting analog-to-digital converters (ADCs) [31, 32]. Both of these circuits are quite suitable for implementation using the eSFQ approach, as they are clocked and therefore naturally coupled to an SFQ clock distribution network.

Circuit design and analysis of performance metrics were achieved with a pre-release version of the *NioPulse* software suite [33], whereas *LASI 7* software [34] was employed for cell and chip layout. Circuit extraction and verification were done with the *InductEx* package [35].

*3.1 Design of eSFQ Shift Registers*
Shift registers are the most natural RSFQ circuits as RSFQ is a sequential logic rather than a combinational logic such as CMOS. A number of RSFQ shift registers designs were developed in the past [36-44]. The most robust and widespread design consists of D flip-flops with an integrated SFQ clock network [39], which we chose for implementation in eSFQ logic. Two different eSFQ versions of these shift registers were implemented: a straight-forward conversion from RSFQ termed "eSR" and a version with an additional magnetic flux bias, "MeSR". The MeSR design is shown to have higher margins retaining the high speed conventional RSFQ design. It also has the potential to achieve lower bias current and thus, considering (3), lower dynamic power dissipation.

*3.1.1. eSFQ shift register (eSR).*
The eSFQ shift register cell, eSR, is depicted in figure 5(a). Its topology can be partitioned into two sections: *clock* and *data*. Junctions $J_{c1}$, $J_{c2}$ and the DMP ($J_1$, $J_2$) make up the clock section, which transmits the clock and interrogates the DMP. Note that we used a counter-flow clocking scheme, as this has generally yielded higher margins in conventional RSFQ designs [39]. If $J_1$ switches, the clock pulse is simply transmitted from *CIn* to *COut*. If $J_2$ switches instead, the clock pulse is transmitted and an output pulse is generated, which exits at *DOut*. In either case, the phase $\phi_x$ increases by $2\pi$. The only bias current injection point for eSR is at point $x$. Inductors $L_2, L_3$ and $L_6$ determine, to a large extent, the bias current distribution between $J_{c1}$, $J_{c2}$ and the DMP (somewhat skewed by Josephson inductances and parasitics). Junctions $J_d$ and $J_2$ make up the data section. After bias current ramp-up, $J_2$ is biased, whereas $J_d$ is not, which corresponds to the cell storing a "1." After the first clock signal, the bias current redistributes to $J_d$, which corresponds to the cell storing a "0". A pulse appears at the output, representing the initially stored "1" as depicted in figure 5(b). For limiting junctions, we chose to use critical damping ($\beta_c = 1$) for convenience of simulation and to prevent undesired junction switching or longer settling time.

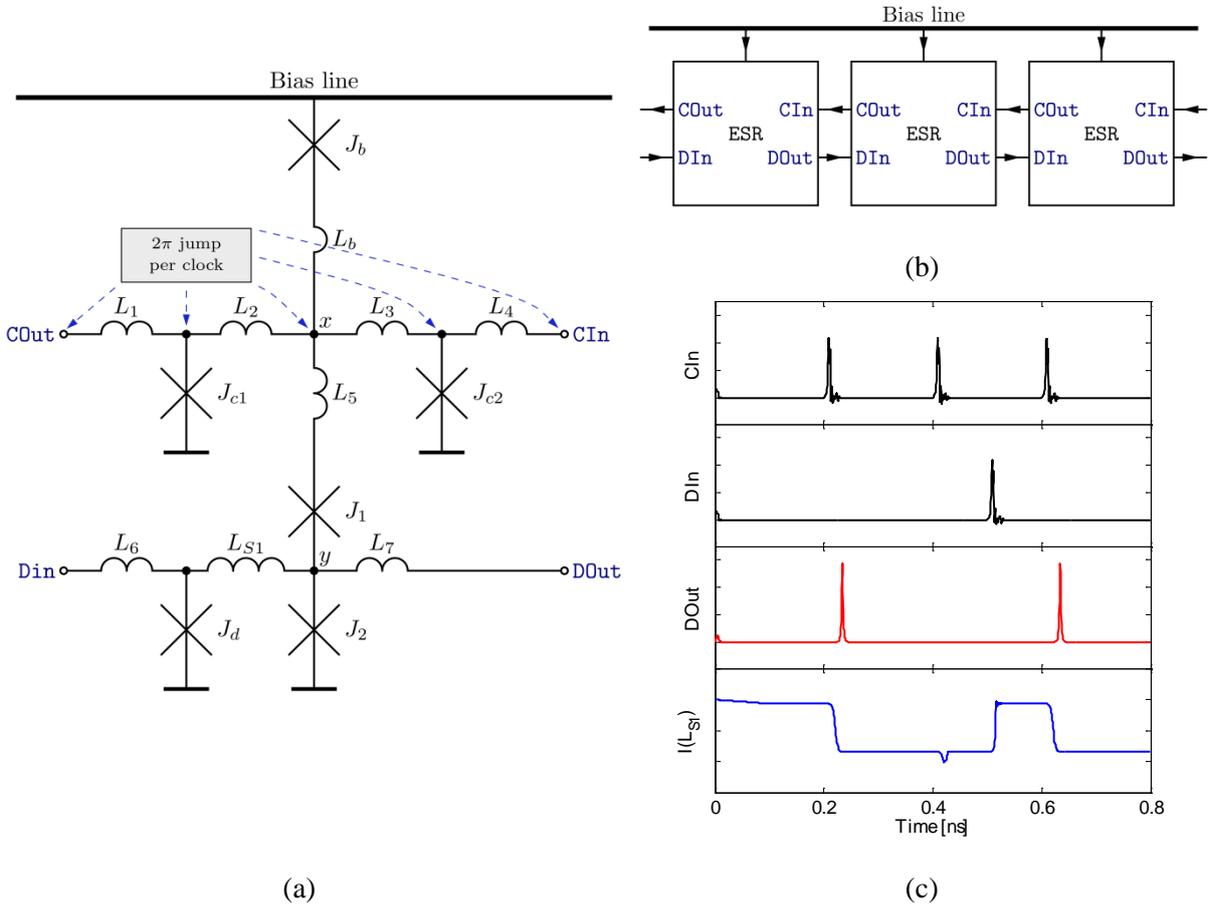

(a)  (b)  (c)

**Figure 5.** eSR - eSFQ shift register cell: Schematic (parasitics are omitted) (a), typical configuration illustrating the counterflow clock (b) and simulated cell operation (c). Circuit parameters for (a): Extracted inductances: L1: 2 pH, L2: 1.9 pH, L3: 1.2 pH, L4: 2.1 pH, L5: 2.1 pH, L6: 3.5 pH, L7: 3.4 pH, LS1: 10.8 pH, Lb: 10 pH. Nominal critical currents: Jc1: 213 µA, Jc2: 250 µA, J1: 313 µA, J2: 188 µA, Jd: 225 µA, Jb: 575 µA. All junctions except J1 are critically damped ($\beta_c = 1$). J1 has $\beta_c \approx 0.25$. Nominal bias per cell: 488 µA.

Margins of operation of the circuit were determined for a 4-bit shift register configuration. The critical parameter was identified as the critical current of junction $J_1$, the upper (escape) junction of the DMP.

One of the reasons for this is the injection of bias current through the DMP as required in accordance with the eSFQ biasing scheme. The difference between the biased and unbiased DMP is evident from the phases of the DMP junctions shown in figure 6.

The grounded junction in a DMP ($J_2$) switches when it is biased, whereas the escape junction ($J_1$) switches when the grounded junction is not biased. The escape junction $J_1$ is, conventionally, not biased. Hence, when $J_2$ is unbiased as it is in case of RSFQ (ERSFQ) circuits, both junctions have a phase near zero, and as $J_1$ has a lower critical current, and is closer to the source of the interrogating pulse, it switches in the presence of an interrogating pulse. When $J_2$ is biased, its phase is nearly critical, with the phase of $J_1$ remaining near zero, making $J_2$ the switching junction when the DMP is interrogated. When biasing through the DMP as in eSFQ, the escape junction $J_1$ is permanently biased. This fixes its operating point phase at greater than zero, increasing its affinity to switch, particularly as it is closer to the source of the interrogating pulse (the Clock node). This reduces the difference between the steady-state phases of $J_1$ and $J_2$ when $J_2$ is biased. In this case, the increased switching affinity of $J_1$ is undesirable (as $J_2$ should then switch) and results in lower parameter margins for $J_1$.

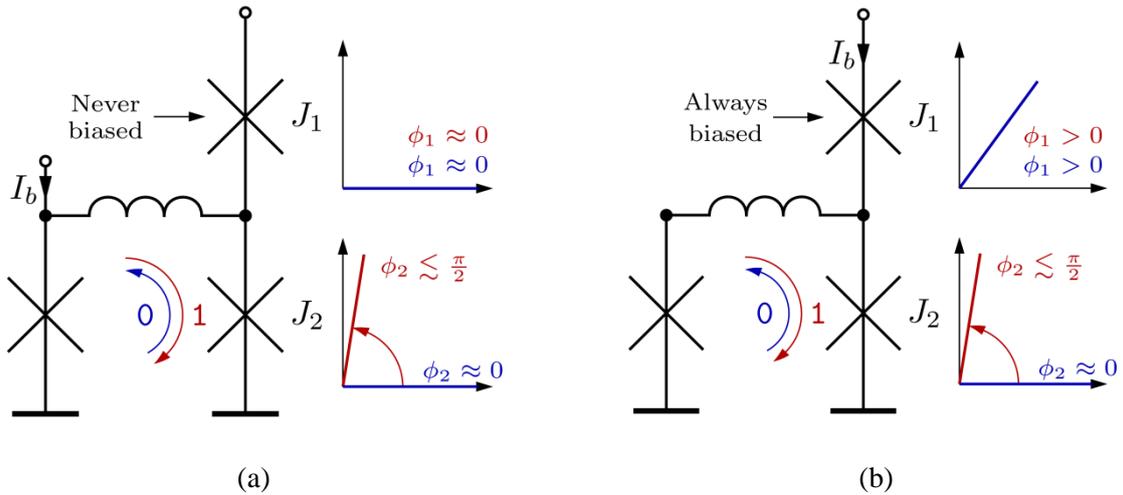

(a)            (b)

**Figure 6.** Comparison of RSFQ (ERSFQ) and eSFQ biasing: Injecting bias current in the conventional DFF leaves the upper junction of the DMP unbiased (a). Moving the bias current injection point to the DMP forces the phase in both DMP junctions in the same direction during ramp-up (b). (Colour may appear only in the online journal.)

In order to improve the margins of $J_1$, it was designed in an overdamped configuration with $\beta_c \sim 0.25$. An overdamped junction exhibits lower switching speeds and is thus less likely to switch before $J_2$ when the DMP receives the interrogating pulse. With the overdamped $J_1$, a 4-bit configuration of eSR achieved critical margins of $\pm 24\%$ and bias margins of $\pm 34\%$, with bias margin relating to the bias of the entire 4-bit test structure and critical parameter the area of $J_1$ for all four eSR cells.

*3.1.2. Magnetically biased eSFQ shift register (MeSR)*
Although overdamping $J_1$ achieves the goal of increased parameter margins, it has the undesired side-effect of increasing data-dependent clock skew. Since $J_1$ switches slower than $J_2$, the clock propagates through the shift register faster when the stored bits are primarily "1"s, and slower if the stored bits are primarily "0"s. This effect is seen also in conventionally biased shift registers, although to lesser extent. In conventionally biased shift registers, this effect is due to an underbiased escape junction $J_1$.

In eSR, it is due to the slow-down imposed for eSFQ biasing. Compared to critical damping, for $\beta_c \sim 0.25$ as used above, the characteristic time of $J_1$ is doubled, potentially halving the maximum clock frequency achievable.

Hence it was desirable to have all junctions equally shunted with $\beta_c \sim 1.0$ and therefore, having the same junction speed, achieving the maximum frequency for a typical 4.5 kA/cm$^2$ critical current density. To accomplish this without punishingly narrow parameter margins, one might investigate several options. For correct operation an interrogating pulse must not cause $J_1$ to switch when $J_2$ is biased. At first glance, keeping the critical current of $J_2$ low should achieve this. Considering figure 5, when the bias current enters at node $x$, some travels down the DMP, biasing junction $J_1$. After crossing $J_1$, the bias current divides again at node $y$. The alternate path to ground through $J_d$ means that some of the bias current leaks away from $J_2$, ensuring that $J_2$ always receives less bias current than $J_1$. Lowering the critical current of $J_2$ increases its Josephson inductance, which exacerbates the leakage effect.

A magnetically introduced corrective flux bias was used to solve the leakage problem, resulting in cell MeSR, depicted in figure 7. The dc flux bias, introduced through $L_f$, forces the current in the storage loop to redistribute as intended, opposing the leakage effect. In this way, the phase offset of $J_2$ (as a result of the eSFQ bias current) can be modified. During circuit optimization, it became apparent that using the flux bias to redirect initial bias current from $J_2$ to $J_d$ was most effective at maximizing parameter margins. A potential advantage of this is that shift registers based on MeSR initially store a "0" which aligns well with conventional RSFQ shift registers.

A further advantage of the flux bias manifests itself in reduced bias current requirements in terms of injected bias current (and corresponding decrease in dynamic power dissipation). As undesired leakage can be avoided and the desired balance in the storage loop be established using the flux bias, less bias current needs to be injected at the injection point $x$. Initial "0" storage and reduced bias current requirements are shown in figure 8. Note that one flux bias line is required to bias the entire shift register, irrespective of its length. For a 4-bit MeSR-based shift register, a critical margin of $\pm 27\%$, and bias margin of $\pm 36\%$ were achieved.

The presence of the additional flux bias line might appear as a complication. In reality, a constant flux bias can be implemented in a variety of ways ranging from a small superconducting loop with frozen-in SFQ to a $\pi$-junction implemented using superconducting-ferromagnetic-superconducting (SFS) Josephson junctions [45-49]. Even for conventional RSFQ circuits, the improved operational margins, bit-error rates, and gate memory non-volatility were reported [50-53].

The reduction of bias current directly translates into the reduction of dynamic power dissipation as $P_d = I_b \Phi_0 f$. This makes the magnetic bias approach especially valuable. As magnetic flux bias is a passive non-switching element, it does not contribute to power dissipation. As is evident from the results of simulations for 20 GHz clock, the eSR shift register consumes $\sim 1.0$ aJ/bit, while the MeSR shift register consumes $\sim 0.8$ aJ/bit. These energies correspond to the centre of the bias current operational region. At the lower limit, the energy per bit operation reaches $\sim 0.5$ aJ/bit.

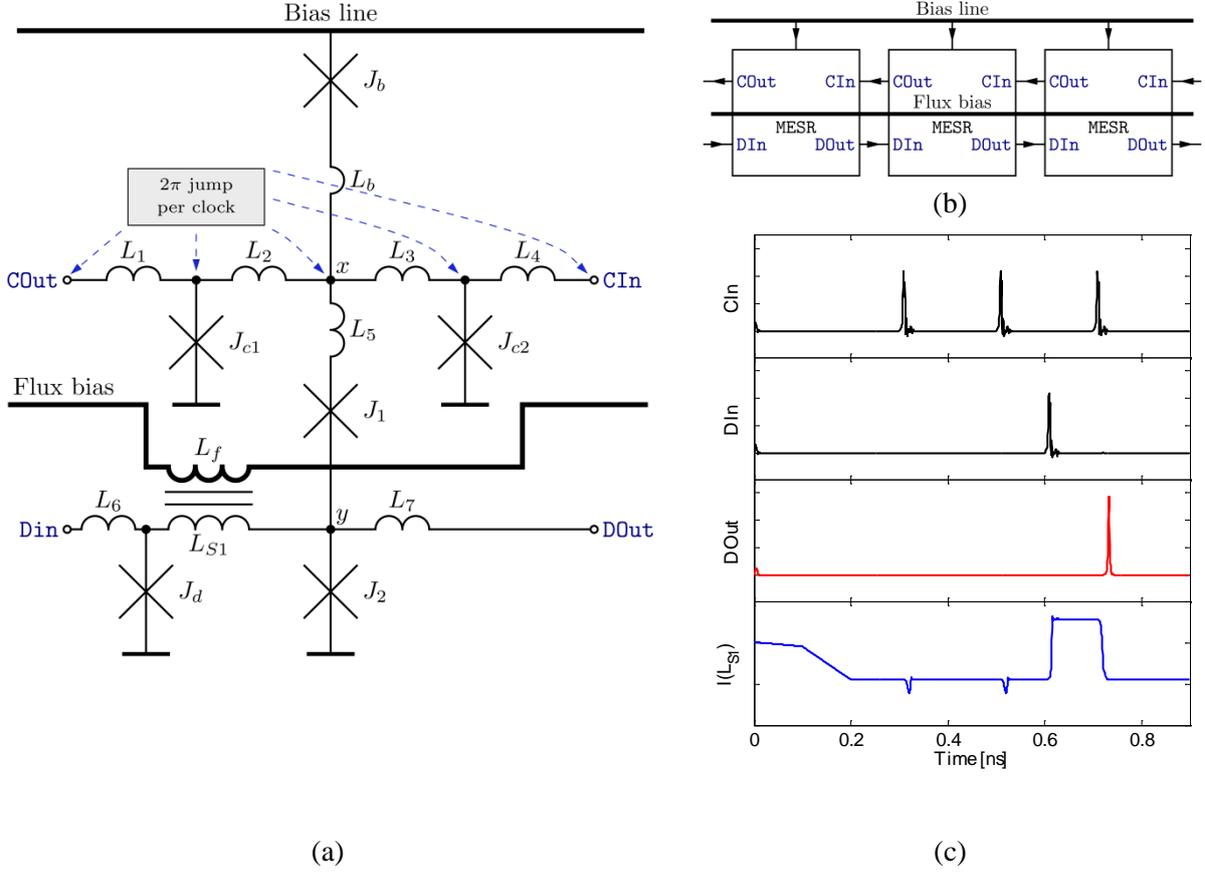

(a) (c)

**Figure 7.** MeSR - eSFQ shift register cell with magnetic flux bias: Schematic of eSFQ shift register cell with flux bias (parasitics are omitted) (a); a typical configuration illustrating the counterflow clocking scheme (b); simulated operation (c). The flux bias ramps up from 0.1ns to 0.2ns, clearly evident in the current trace for $L_{S1}$. After flux-bias ramp-up, MeSR is non-storing (cf. Figure 5). Circuit parameters for (a): Extracted inductances: L1: 2.2 pH, L2: 1 pH, L3: 1.8 pH, L4: 2.0 pH, L5: 1.8 pH, L6: 2.8 pH, L7: 4.2 pH, LS1: 11.2 pH, Lf: 11.5 pH, k: 0.25, Lb: 10 pH. Nominal critical currents: Jc1: 188 µA, Jc2: 188 µA, J1: 288 µA, J2: 200 µA, Jd: 163 µA, Jb: 525 µA. All junctions are critically shunted. Nominal bias per cell: 400 µA. Nominal flux bias: 400 µA.

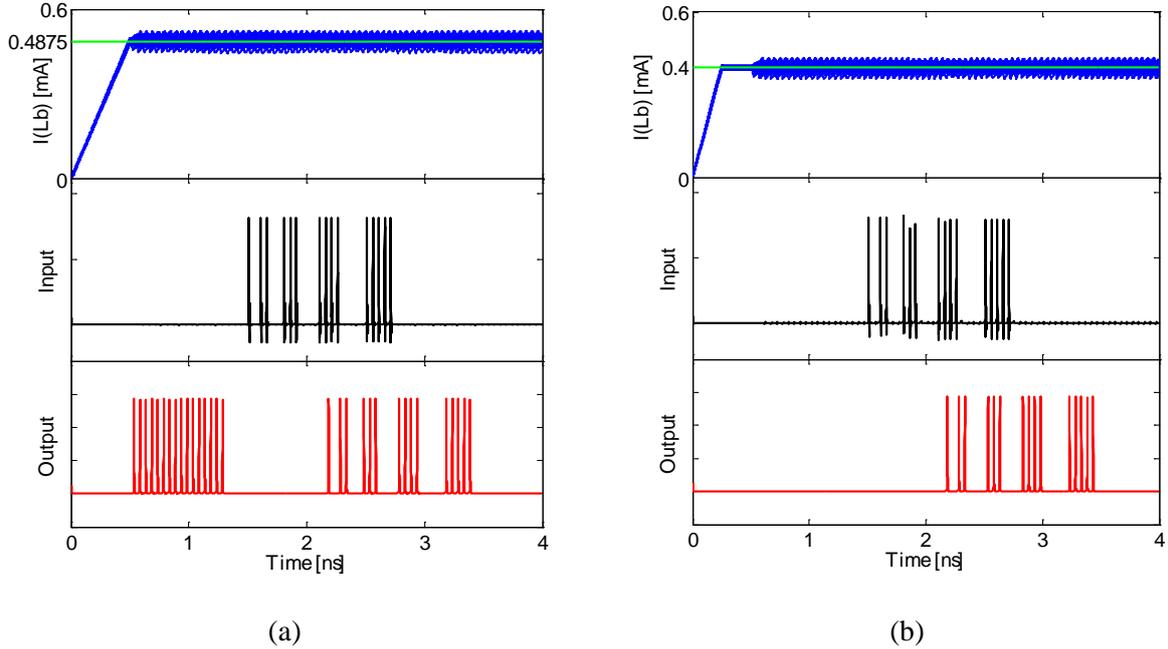

(a)                    (b)

**Figure 8**. Results of simulation with a 20 GHz clock for 1.0 aJ/bit eSR- (a) and 0.8 aJ/bit MeSR-based (b) 16-bit eSFQ shift registers at bias current corresponding to centre of operational region. Bias currents distribute correctly, with acceptable distortion through switching events. For the eSR-based shift register, 16 output bits are immediately observed after starting the clock. In both cases, the input pattern is reproduced at the output. Lower bias current requirements are evident for the magnetic flux-biased shift register.

*3.1.3. eSFQ deserializer – eDES*

RSFQ deserializers (demultiplexers) generally follow two different approaches: a binary tree [30, 37, 54-58] or a shift-and-dump [59-61] architecture. For conversion to eSFQ, we chose the latter approach as it has found more applications in practical circuits due to its high modularity and simple timing. Our eSFQ deserializer is based on a dual-port D flip-flop or $D^2$-cell [62], which is a derivation of the B flip-flop [63]. One port is intended for serial shifting of data, the other for parallel readout. An $n$-bit deserializer divides a serial stream of bits into $n$ parallel streams.

The designed eSFQ deserializer cell, eDES, is depicted in figure 9. The two readout ports are topologically symmetrical, both achieving destructive readout of stored flux. Note the additional escape junction in each readout arm ($J_{der}$, $J_{des}$). The deserializer cell contains two DMPs, suggesting two bias injection points. As in MeSR, a flux bias is employed in the data section to achieve the desired bias current distribution between $J_d$, $J_{2r}$, $J_{2s}$. All junctions were designed to be critically shunted. When correctly biased, eDES stores a "0" after ramp-up.

There are essentially two clocks that thread each deserializer cell. The symmetry of the cell and size of the limiting junction means that the per-bit switching energy required by the shift operation as well as the read operation is comparable to that of the MeSR-based shift register. For normal operation the ratio of the clock frequencies depends on the length of the deserialiser. A per-bit switching energy is thus not meaningfully ascribed to the deserializer cell, but for long deserializers the per-bit switching energy of the deserializer approaches that of the MeSR-based shift register.

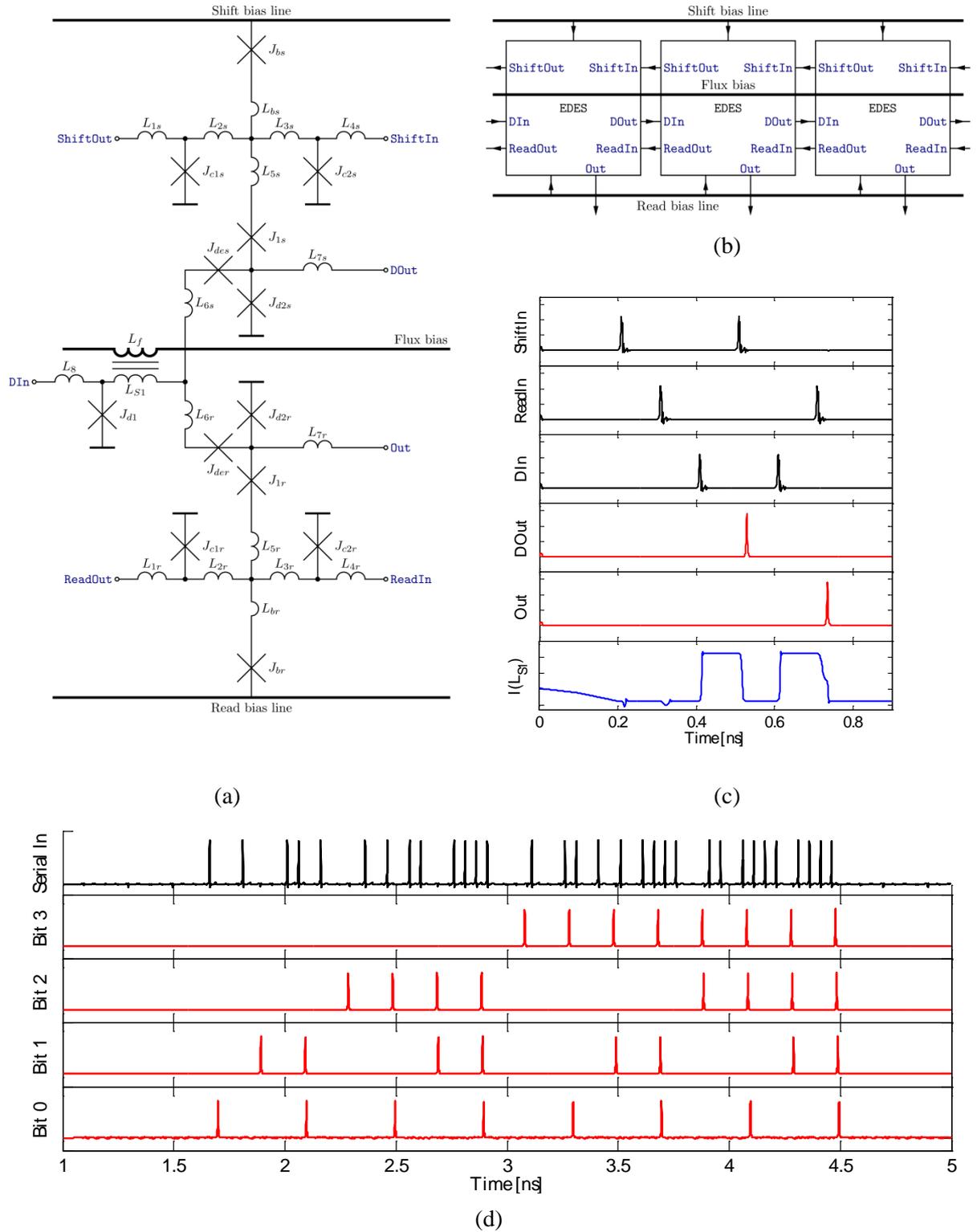

**Figure 9.** eDES - eSFQ deserializer cell with magnetic flux bias. Schematic of eDES cell (a); a deserializer configuration (b); the operation of eDES (c). Flux bias is ramped up from 0.1 ns to 0.2 ns, as evident in the bottom trace. Initially, eDES is non-storing, but when a pulse arrives at DIn, it is stored, readable by both a Shift or a Read pulse. Simulated 4-bit operation of the deserializer at 20 GHz (d). Clearly, the input signal (a count from 0 to 15) is parallelized, resulting in 4 output streams. Circuit parameters for (a): Extracted inductances: $L_{1s}$: 2.9 pH, $L_{2s}$: 2.0 pH, $L_{3s}$: 1.4 pH, $L_{4s}$: 1.2 pH, $L_{5s}$: 0.2 pH, $L_{6s}$: 0.9 pH, $L_{7s}$: 4.2 pH, $L_8$: 2.0 pH, $L_{S1}$: 3.1 pH, $L_{bs}$: 10 pH, $L_f$: 16.1 pH, k: 0.24, $L_{1r}$: 1.7 pH, $L_{2r}$: 1.7 pH, $L_{3r}$: 1.6 pH, $L_{4r}$: 2.5 pH, $L_{5r}$: 0.2 pH, $L_{6r}$: 0.6 pH, $L_{7r}$: 3.4 pH, $L_{br}$: 10 pH. Nominal critical currents: $J_{c1s}$: 163 µA, $J_{c2s}$: 188 µA, $J_{1s}$: 288 µA, $J_{d2s}$: 188 µA, $J_{des}$: 200 µA, $J_{d1}$: 150 µA, $J_{bs}$: 500 µA, $J_{c1r}$: 188 µA, $J_{c2r}$: 188 µA, $J_{1r}$: 225 µA, $J_{d2r}$: 188 µA, $J_{der}$: 138 µA, $J_{br}$: 500 µA. All junctions are critically shunted. Nominal bias per cell: 413 µA (Clock), 413 µA (Read), nominal flux bias: 500 µA.

## 4. Experimental Evaluation
### 4.1. Sample Layout and Fabrication

In order to investigate eSFQ logic experimentally, the eSR, MeSR and eDES cells were laid out and their circuits were reoptimized to account for the extracted layout parasitics. Several eSFQ shift registers with 16- and 32-bit length, as well as deserializers with 4-, 8- and 16-bit lengths were assembled. Figure 10 shows examples of the experimental eSFQ chips designed for fabrication using the HYPRES Niobium superconductor integrated circuit fabrication process [64-66]. To investigate the performance of the designed eSFQ circuits in a variety of environments, 12 test structures were laid out across five 5 x 5 mm$^2$ chips for the fabrication with a 4.5 kA/cm$^2$ Josephson junction critical current density.

Figure 11 shows examples of the fabricated eSR, MeSR circuits. As is evident, they differ in the escape junction $J_1$ shunt resistor in order to achieve overdamping for the eSR design and critical damping for the MeSR design. For MeSR, the magnetic flux bias was implemented as a superconducting line under the cell storage inductors to induce magnetically the required phase shift. Microphotographs of the deserializer are depicted in figure 12. In order to protect circuits from flux trapping, ground plane moats [67] were employed as well as ground plane holes covering unused chips areas.

To concentrate design effort on the eSFQ demonstrator cells and to minimize the probability of failure in the periphery circuits, existing conventional RSFQ cells from the HYPRES cell library were employed as a testbed. These comprise standard interfaces to room-temperature circuitry, such as dc/SFQ and toggle-type SFQ/dc converters [28, 30]. Figure 12(b) shows an RSFQ testbed made of these standard library RSFQ cells. The test chips also contain standard diagnostic circuits for fabrication process control visible in figure 10.

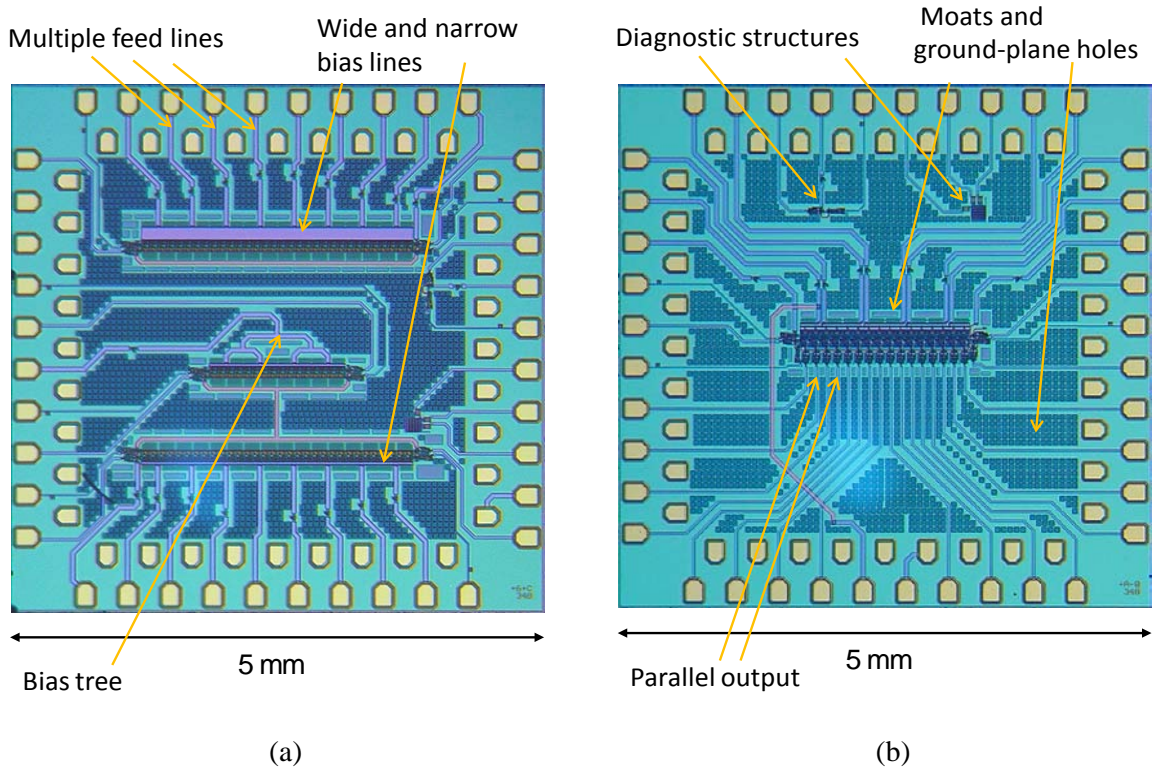

**Figure 10.** Layouts of two of the five 5x5 mm² ICs fabricated with HYPRES's 4.5 kA/cm² process, one with three shift registers (a), and one with a 16-bit deserializer (b). Each bias line is fed from multiple contact pads, which enables experimental investigation of different bias fan-in configurations.

Besides establishing functional correctness of the designed eSFQ cells, the objective was to investigate experimentally the effects of the bias current distribution in superconducting biasing network at the initial bias current ramp-up. Since eSFQ circuits do not have resistors in the bias network, the bias distribution relies on interplay between specific inductances of the bias lines and gate current limiting junctions [21]. This is not easy to simulate as the circuit initialization is inherently a slower process than its SFQ operation. For this reason, several versions of shift registers with different width (specific inductance) of bias distribution buses and different current injection fan-in were designed.

Thus, in addition to laying out different combinations of the basic designed cells, these were placed in a variety of different bias lines. The designed bias lines are characterized by the cell-to-cell inductance of the line, $L_q$, as well as the line-to-cell limiting inductance, $L_b$. Bias lines of two different widths (narrow: $L_q \approx 1.5\text{pH}$, wide: $L_q \approx 0.5\text{pH}$) were laid out and combined with three lengths of limiting inductor (short: $L_b \approx 10\text{pH}$, medium: $L_b \approx 50\text{pH}$, long: $L_b \approx 150\text{pH}$).

For each laid-out structure, bias fan-in was conservatively high. Each structure has its own bias line, which is shared by all cells in the structure (deserializer structures each have two bias lines, one for the read- and one for the shift operation). One bias pin was allocated to every four cells in a structure. This enables comprehensive investigation of different bias-current fan-in configurations (biasing from all available pins or biasing from one pin only, for example). Figure 10 depicts two examples of chip layout, illustrating the high number of bias-current pins. To further the breadth of this investigation, one structure was equipped with a bias-current divider that binds the four bias line entry points of the 16-bit shift register to a single pin.

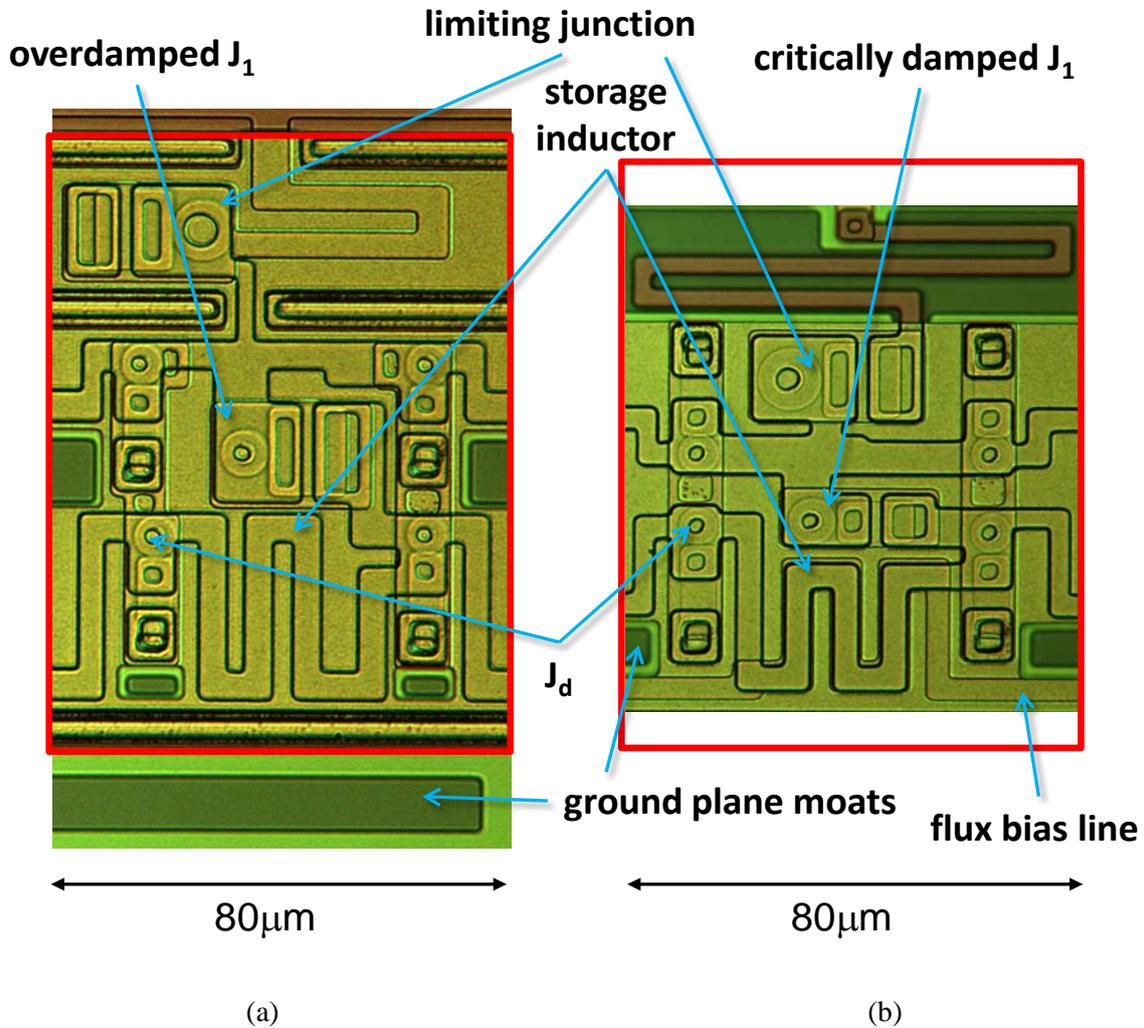

**Figure 11.** Layouts of eSFQ shift register cells eSR with overdamped $J_1$ (a) and MeSR with critically damped $J_1$ and a flux bias line inductively coupled to cell storage inductor (b). Cell sizes (indicated by red boundary): eSR: 80x110 µm², MeSR: 80x105 µm². These dimensions do not include the bias line.

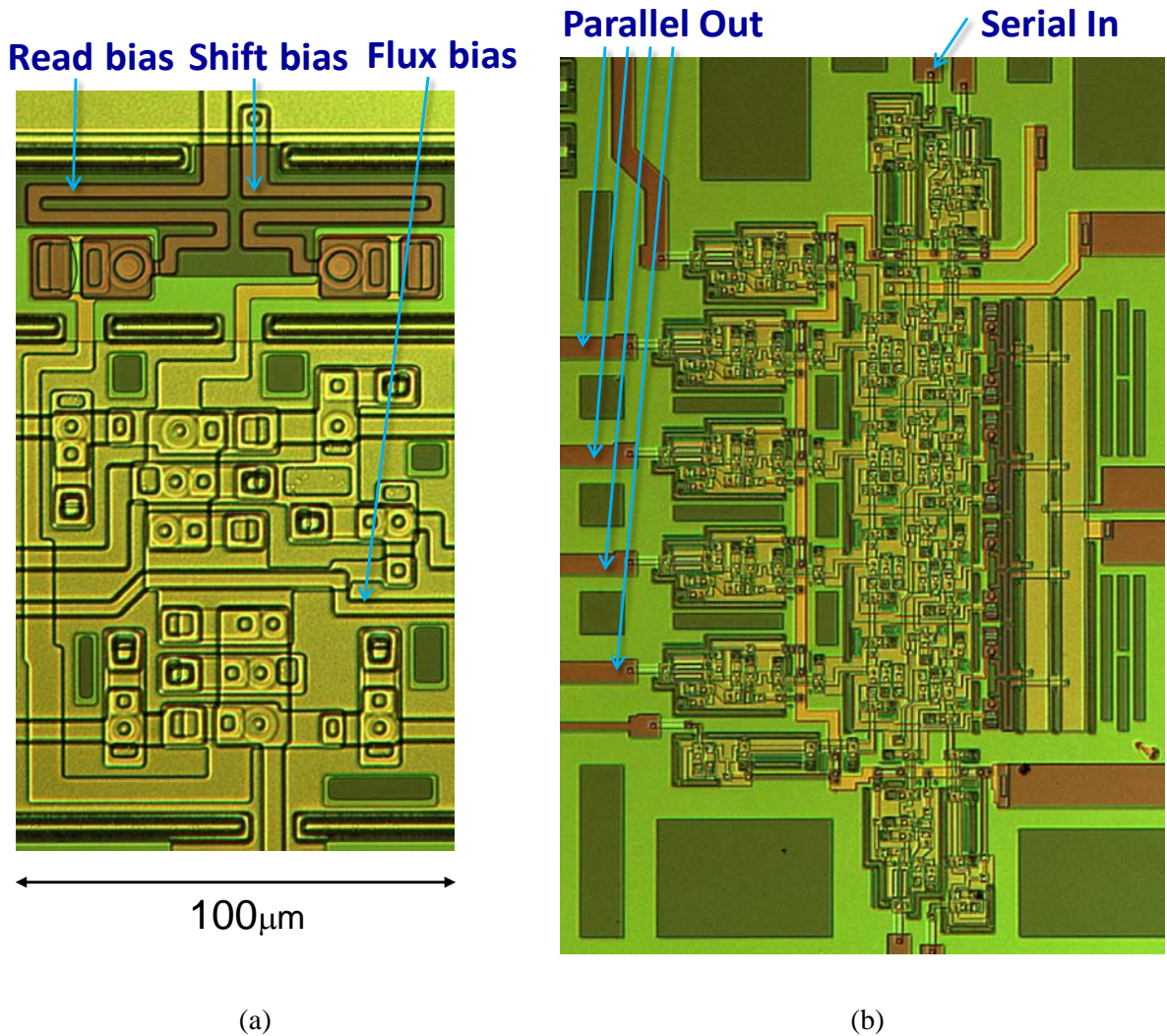

(a)                              (b)

**Figure 12.** Photographs of deserializer test structures: the deserializer base cell eDES (a), and a corresponding 4-bit eSFQ deserializer (b). Bias lines and some key devices are indicated. A set of dc/SFQ and SFQ/dc converters enable the interface with room temperature electronics.

*4.2. Test Results*
Experimental evaluation was performed with test patterns applied and responses measured with the Octopux system [68]. Each chip was tested in a liquid helium dewar using HYPRES standard cryoprobes. Correct operation of the shift register structures was established by feeding in a bit pattern and verifying its transmission with the correct delay (in terms of clock events). Correct $n$-bit deserializer operation was established by feeding in a pattern of length $n$ with the shift clock, applying a read pulse, and then verifying the parallel readout against the input pattern. This process was repeated several times to verify deserializer operation. Figures 13, 14 depict examples of the measured correct test patterns of the 16-bit eSFQ shift registers and deserializers. For exhaustive testing, not only uniform clock and data patterns were employed, but also randomly generated ones.

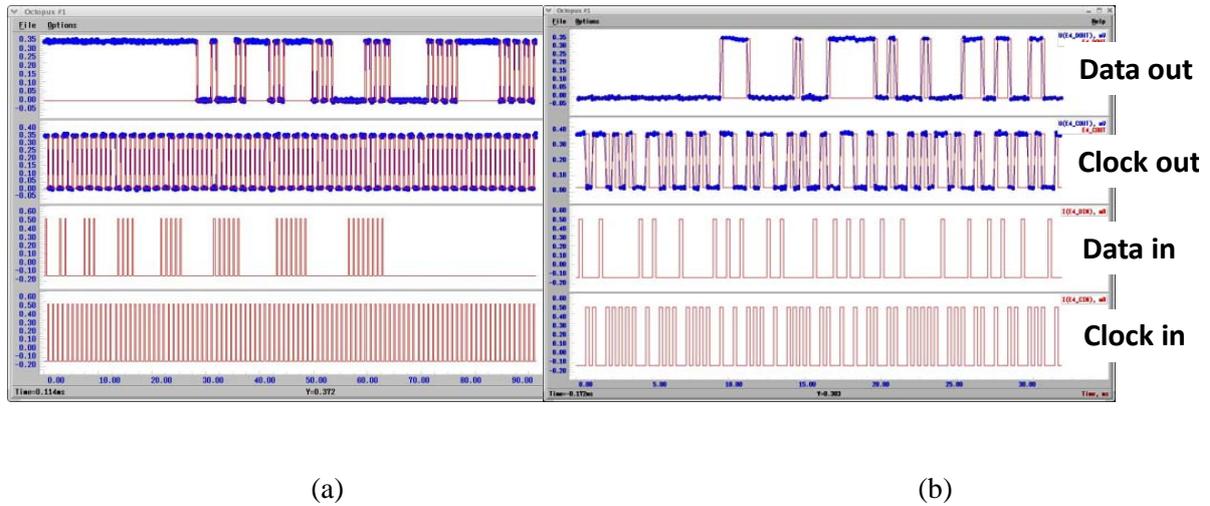

(a)                          (b)

**Figure 13.** Measured correct low-speed functionality of 16-bit eSFQ shift registers: simple pattern (a), randomly generated data and clock pattern (b). All 9 shift registers were fully operational.

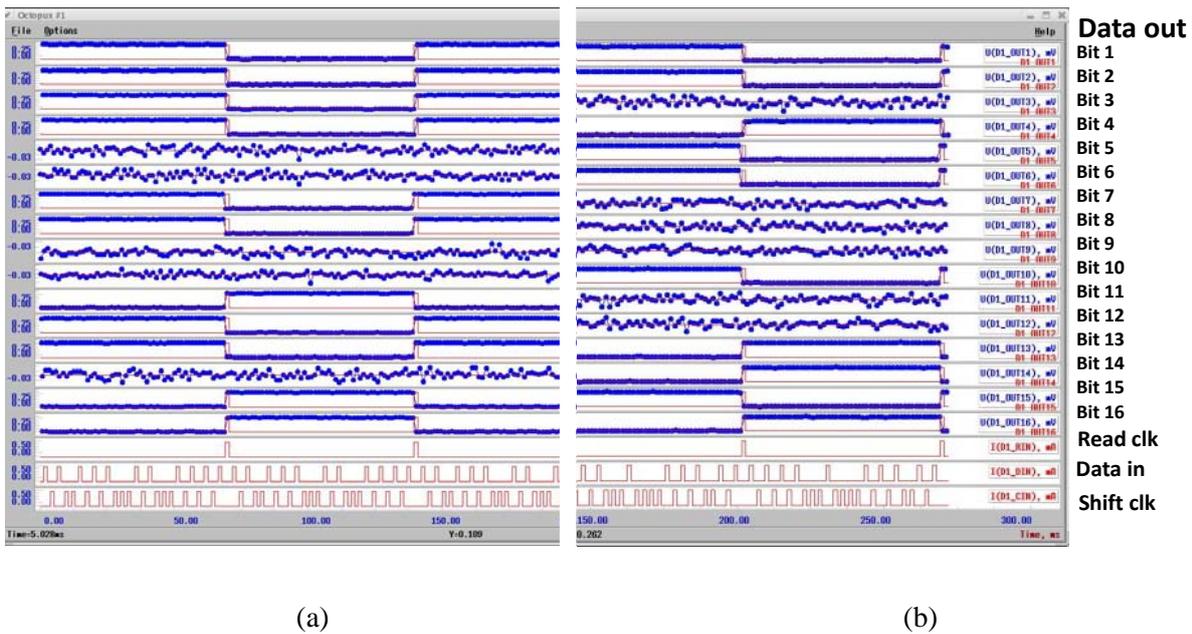

(a)                          (b)

**Figure 14.** Measured correct low-speed functionality of 16-bit deserializer for different input patterns. This circuit divides the input serial stream into 16 output parallel streams. Random data and shift clock pulses are used for testing. Note that measured voltages were scaled automatically, so that constant-level logic signals such as Bit 7 in (b) appear noisy.

To determine the bias margins of investigated structures, random 200-bit test patterns were applied to the devices under test for various bias currents. Various bias-current feeding configurations were investigated, most notably biasing from one pin only (repeated for each available pin), and biasing from all pins simultaneously. Only the largest identified continuous region of operation was considered. All tested structures passed functional testing for all tested patterns. The measured results of the bias current margin investigation are listed in table 1.

**Table 1.** Experimentally determined bias margins for eSFQ test structures across five chips: a comprehensive set of devices in different bias configurations, measured to establish functional correctness of the devices and attempt to identify desirable traits of the bias line layout.

| Kind | Len | Bias Line | | Comment | Bias margins [mA] | |
| --- | --- | --- | --- | --- | --- | --- |
| | | $L_q$ | $L_b$ | | All pins | Single pin |
| Shift Reg | 32 | Narrow | Medium | | 13.9 ± 27.5% | 15.4 ± 14.5% |
| Shift Reg | 16 | Narrow | Medium | Bias Tree | - | 6.71 ± 35.5% |
| Shift Reg | 32 | Wide | Medium | | 14.0 ± 29.8% | 13.7 ± 24.5% |
| Shift Reg | 16 | Narrow | Short | | 8.38 ± 21.8% | 8.2 ± 24.6% |
| Shift Reg | 16 | Narrow | Medium | | 7.0 ± 30.8% | 7.7 ± 21.6% |
| Shift Reg | 16 | Narrow | Long | | 7.47 ± 23.1% | 7.9 ± 18.8% |
| Shift Reg | 32 | Narrow | Medium | Flux Bias | 13.0 ± 27.0% | Not tested |
| Shift Reg | 16 | Narrow | Medium | Flux Bias | 7.6 ± 13.0% | Not tested |
| Shift Reg | 16 | Wide | Medium | Flux Bias | 8.4 ± 19.7% | Not tested |
| Deserialiser | 16 | Narrow | Medium | Flux Bias | 6.8 ± 14.4% (Clk) 7.7 ± 12.3% (Rd) | Not tested |
| Deserialiser | 8 | Narrow | Medium | Flux Bias | 3.7 ± 22.9% (Clk) 3.3 ± 31.9% (Rd) | Not tested |
| Deserialiser | 4 | Narrow | Medium | Flux Bias | 1.4 ± 47.4% (Clk) 1.5 ± 36.5% (Rd) | Not tested |

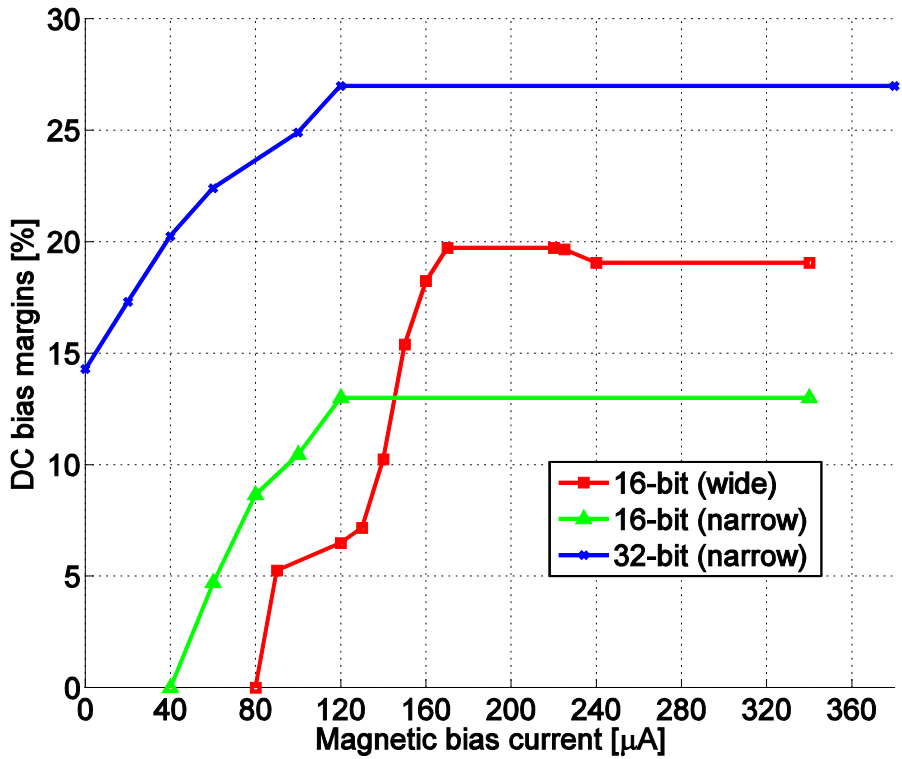

**Figure 15.** Measured bias margins vs. magnetic bias for different MeSR-based shift registers.

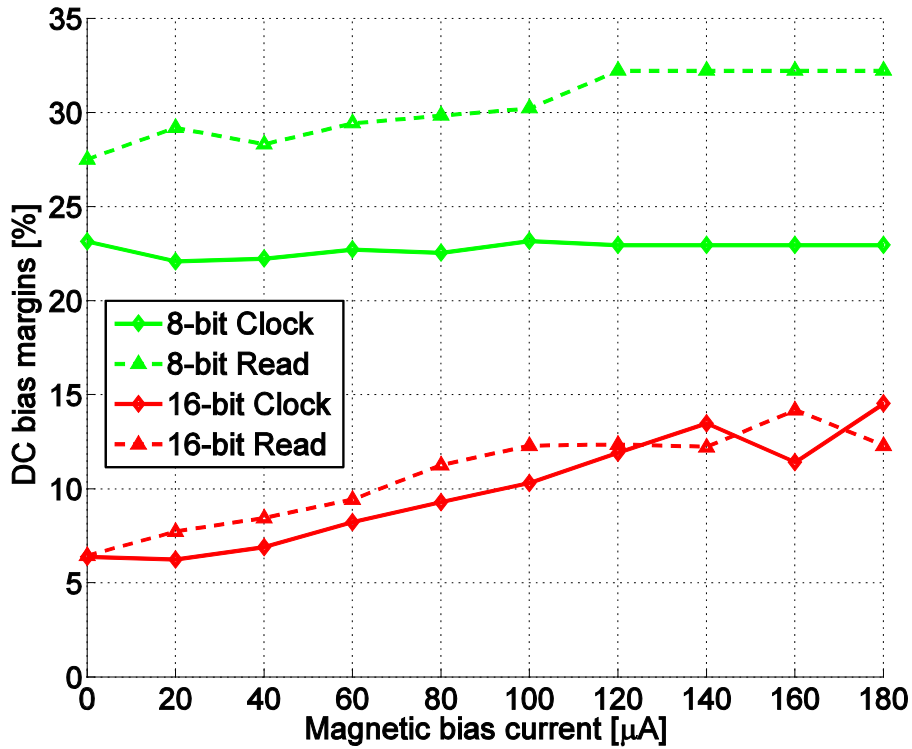

**Figure 16.** Measured bias margins vs. supplied magnetic bias current of 8-bit and 16-bit deserializer test structures.

The measured bias margins roughly conform to expectations extrapolated from simulated results. Simulations relied on (small) 4-bit configurations to reduce computation time to design-friendly speeds and were performed in high-speed testbeds, not reflecting the actual devices under test or their periphery. The observed agreement with simulations confirmed that the designed structures are robust and scale well.

As figure 15 indicates, the MeSR-based shift registers (with magnetic flux bias) functioned only when the magnetic bias is applied, which corresponds to our simulations. Bias margins do not seem dependent on the length of the shift register structure, although the dataset is too small to identify definite trends. As evident from figure 15, the margins of operation for a 32-bit shift register are better than for 16-bit circuit. Although, we did not test many copies of the same circuits to confirm this, we attribute the measured data to lesser influence of adjacent RSFQ test input/output to biasing of the ESFQ circuit under test.

The deserializer structures under test worked without the magnetic flux bias applied. As demonstrated in figure 16, applying the magnetic flux bias improves their bias margins, doubling them in the case of the 16-bit deserializer. For the 16-bit deserializer, the clock and read bias margins exhibit comparable absolute values and dependence on the magnetic flux bias, which is consistent with simulations. The 8-bit structure does not mirror this symmetry, which may indicate higher susceptibility to the periphery.

Comparatively low bias margins were recorded for the 16-bit deserializer. We attribute this to the relative complexity of the test structure and to an ill-designed monitor setup. The high fan-out of the deserializer requires a large number of SFQ/dc monitors, as well as several dc/SFQ converters to apply the test patterns. Due to the high pin count of the test structure (two bias lines, parallel output), all peripheral cells were biased from a single pin, which yielded low margins for the peripheral bias, potentially depressing the margins of the device under test.

Simulations did not yield conclusive results for designing optimal bias line inductances and limiting inductors, although minimizing the bias line inductance $L_q$ seemed to be helpful in obtaining better bias-current distributions. Similarly, conclusive results about best practices for bias-current fan-in were not obtained. The test structures, meant to comprehensively test the effect of inductances on margins, did not yield conclusive results either. We conclude from this that these inductances, as well as bias current fan-in, at least when constrained to the parameter ranges investigated, do not predictably affect bias current distribution. As the ratio of bias current to critical current of the limiting junctions $J_{bi}$ is less than 1, there is some wiggle-room in which bias currents may re-distribute themselves between the branches. The exact distribution of the bias currents depends on the nature of the bias current ramp-up curve, how it behaves before settling on its final value $I_b$. As this is largely unpredictable and may be affected by flux quanta trapped in the bias network during cool-down, it may explain the differences in margins measured for similar test structures. Promising are the high bias margins measured for the 16-bit shift register biased from a single pin feeding a bias tree. This suggests that the off-chip biasing effort for eSFQ systems should be comparable to systems based on conventional RSFQ.

## 5. Discussion and Conclusions

We have demonstrated for the first time eSFQ digital circuits – a new ultra-low power RSFQ-type logic capable of achieving a significant increase in energy efficiency of computing systems. The demonstrated eSFQ shift registers reached ~0.8 aJ/bit. This number includes the integrated SFQ clock lines. The achieved energy per bit is over two orders of magnitude better than that of the same circuits implemented in conventional RSFQ logic.

Similar to ERSFQ, another energy efficient RSFQ-type logic, eSFQ relies on limiting Josephson junctions to distribute the dc bias to logic gates. In contrast to ERSFQ, the limiting junctions do not switch during circuit operation and are needed only for the initial bias current ramp-up. This is achieved by the bias current injection via two-junction decision making pairs (DMPs) which have equal phases during the gate operation independent of digital data. This also allows eSFQ circuits to operate without large bias inductances otherwise needed to minimize data dependent bias current fluctuations. As a result, the eSFQ circuit layouts are more dense and easier to scale.

However, we found that the injection of bias current via DMPs depresses parameter margins. In this work, we explored ways to rectify this effect by using either stronger junction damping (slowing down the DMP escape junction) or passive phase shifters. The first method is simpler in the implementation but leads to a reduction of the maximum speed of operation. The second method does not limit the maximum clock frequency but requires an introduction of extra phase shifting elements such as flux bias line or π-junctions.

Passive phase shifters would bring an additional, perhaps even more significant result: a reduction of the required dc bias for eSFQ gates. We demonstrated that phase shifters allow a ~20% gate bias reduction which directly translates to the corresponding ~20% reduction of the gate dynamic power dissipation, $P_d = \Phi_0 I_b f_{CLK}$. We believe that this number can be further improved with the targeted circuit optimization. For example, the critical currents used in these circuits are in a 180-300 μA range, which is larger than required by thermal noise at 4 K.

In contrast to a simple flux biasing line implemented in this work, the introduction of phase shifting π-junctions would require the incorporation of ferromagnetic materials into a conventional superconducting fabrication process. This might look cumbersome and expensive at first, but the work on the superconductor-ferromagnetic fabrication process is already happening. It is motivated by the recent efforts in superconducting magnetic memory developments and research in superconducting spintronics [69, 70]. We expect that superconductor–ferromagnetic phase shifters will be preferable for eSFQ circuits.

The demonstrated eSFQ shift register design is quite compact comprising five junctions per bit excluding the passive bias-limiting junction. The total number of junctions for a 16-bit shift register is 80. This compares favourably to an RQL shift register with 8 junctions per bit [19] and low-power RSFQ shift register [18] with 200 junctions for an 8-bit circuit.

While achieving a significant improvement in power dissipation, the demonstrated eSFQ logic retains all key advantages of conventional RSFQ logic: high speed, high throughput, dc bias, controllable and programmable SFQ clock, and lossless interconnects. As opposed to ac biasing and global clock, the eSFQ dc bias and locally controllable SFQ clock is particularly advantageous for scaling up integrated circuit complexity to millions of junctions. Simple eSFQ layout requirements without the need for transformers and microwave plumbing also bode well in terms of scaling up the circuit density. Finally, the ability to control the SFQ clock distribution allows management of eSFQ circuit power dissipation. This is the pre-requisite for the development of energy proportional processors – the ultimate goal of computing system developers [1, 5].

Improving energy efficiency of microprocessors for high-end computing systems will not be complete without addressing random access memory (RAM) capable of matching in speed and power the energy-efficient digital circuits described above. Recently, an approach based on the use of superconducting–ferromagnetic structures was proposed to construct an energy-efficient RAM compatible with eSFQ logic [69, 70].

Superconducting systems require cryocooling. The efficiency of 4 K cryocoolers ranges from ~10,000 W/W for small cryocoolers (heat capacity of < 1W) [71] to < 400 W/W for large machines (heat capacity of 600-900 W) [72, 73]. For high-end computing systems, the larger cryocoolers are relevant. For example, for the Linde LR280 with 360 W/W efficiency [72], this puts the eSFQ circuit demonstrated in this work at ~ 290 aJ/bit. This is close to the projected bit energy for a future CMOS gates, however, one should realize that the biggest energy loss (>~pJ/bit) in CMOS circuits is in data movement. In contrast, superconducting SFQ circuits including eSFQ can use ballistic data transport at the similar sub-aJ per bit level as for logic and register circuits. There is also a potential advantage of superconducting SFQ circuits over room-temperature competition in power density, which constrained the progress towards faster CMOS circuits. Assuming that eSFQ circuits can be scaled to the CMOS circuit densities, power density will be at least three orders of magnitude lower, as the cryocooling penalty does not change this difference. The practical advantage of this much lower power density requires further study to account for comparative heat removal capabilities and operation temperature ranges.

**Acknowledgments**
The work is supported in part by US DoD contract W911NF-09-C-0036. The authors wish to thank S. Kaplan for many discussions on RSFQ and eSFQ operation, V. Semenov for useful discussions and NioCAD Pty (Ltd.) for circuit design software and support, A. Kirichenko and A. Inamdar for the contribution of peripheral cells for the IC layouts, as well as I. Vernik for help and verification of circuit layouts. Special thanks go to the HYPRES fabrication team of D. Yohannes, J. Vivalda, R. Hunt, D. Donnelly, D. Amparo and S. Tolpygo who delivered the integrated circuits on time and within specifications. We also thank M. Manheimer and S. Holmes for attention to this work, useful discussions and advice on the manuscript.